\documentclass[aps,prl,amsmath,amssymb,twocolumn,superscriptaddress]{revtex4}
\usepackage{graphicx}
\usepackage{braket}
\usepackage{color}
\usepackage{soul}
\usepackage{dcolumn}  
\usepackage[dvipsnames]{xcolor}

\begin{document}


\title{Single-electron control in a foundry-fabricated two-dimensional qubit array}

\author{Fabio~Ansaloni\textsuperscript{\dag}}
\affiliation{Center for Quantum Devices, Niels Bohr Institute, University of Copenhagen, 2100 Copenhagen, Denmark}

\author{Anasua~Chatterjee\textsuperscript{\dag}}
\affiliation{Center for Quantum Devices, Niels Bohr Institute, University of Copenhagen, 2100 Copenhagen, Denmark}

\author{Heorhii~Bohuslavskyi}
\affiliation{Center for Quantum Devices, Niels Bohr Institute, University of Copenhagen, 2100 Copenhagen, Denmark}

\author{Benoit~Bertrand}
\affiliation{CEA, LETI, Minatec Campus, Grenoble, France}

\author{Louis~Hutin}
\affiliation{CEA, LETI, Minatec Campus, Grenoble, France}

\author{Maud~Vinet}
\affiliation{CEA, LETI, Minatec Campus, Grenoble, France}

\author{Ferdinand~Kuemmeth\textsuperscript{*}}
\affiliation{Center for Quantum Devices, Niels Bohr Institute, University of Copenhagen, 2100 Copenhagen, Denmark}
\affiliation{\textsuperscript{*}kuemmeth@nbi.dk\\\textsuperscript{\dag}These authors contributed equally to this work.}

\date{\today}

\begin{abstract}
	
\end{abstract}

\maketitle

\textbf{
Silicon spin qubits have achieved high-fidelity one- and two-qubit gates \cite{Veldhorst2015,Muhonen2014,Watson2018,He2019, Zajac2018}, above error-correction thresholds \cite{Yoneda2018}, promising an industrial route to fault-tolerant quantum computation.
A significant next step for the development of scalable multi-qubit processors is the operation of foundry-fabricated, extendable two-dimensional (2D) arrays. 
In gallium arsenide, 2D quantum-dot arrays recently allowed coherent spin operations and quantum simulations \cite{Mortemousque2018,Dehollain2019}. 
In silicon, 2D arrays have been limited to transport measurements in the many-electron regime~\cite{Betz2016}. 
Here, we operate a foundry-fabricated silicon 2x2 array in the few-electron regime, achieving single-electron occupation in each of the four gate-defined quantum dots, as well as reconfigurable single, double, and triple dots with tunable tunnel couplings.
Pulsed-gate and gate-reflectometry techniques permit single-electron manipulation and single-shot charge readout, while the two-dimensionality allows the spatial exchange of electron pairs. 
The compact form factor of such arrays, whose foundry fabrication can be extended to larger 2xN arrays, along with the recent demonstration of coherent spin control \cite{Maurand2016} and readout~\cite{Crippa2019,Urdampilleta2019},
paves the way for dense qubit arrays for quantum computation and simulation \cite{Vandersypen2017}.
}

\begin{figure}
	\includegraphics[scale=1]{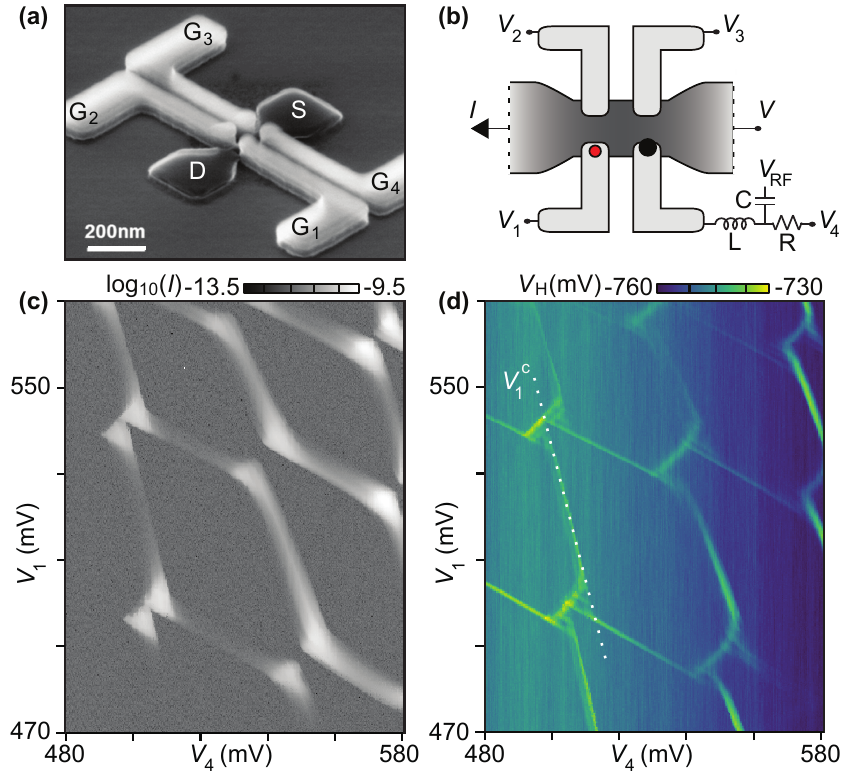}
	\caption{\textbf{Compensated control voltages within a two-dimensional silicon quantum-dot array.}
(a) Foundry-fabricated undoped silicon channel connected to reservoirs (dark grey), with four gate electrodes (light grey). 
This SEM image shows a device from a different fabrication run without back-end~\cite{Barraud2016}. 
(b) Device schematic for the example of a few-electron double dot underneath gates G$_\mathrm{1}$ and G$_\mathrm{4}$, induced by appropriate control voltages $V_\mathrm{1-4}$. 
Each of the three qubit dots (dot 1 indicated in red) capacitively couples to the sensor dot (black), which can be monitored using rf reflectometry off an inductor (L) wirebonded to G$_\mathrm{4}$. 
(c)-(d) Charge stability diagram of the double quantum dot in (b), acquired at fixed source-drain bias $V=-3~$mV. 
Source-drain current $I$ and demodulated reflectometry voltage $V_\mathrm{H}$ measured as a function of $V_\mathrm{1}$ and $V_\mathrm{4}$. 
The dotted white line defines a compensated voltage $V_\mathrm{1}^c$ that controls the chemical potential of dot 1 without affecting the chemical potential of dot 4. Control voltages $V_\mathrm{1,2,3}^c$ for other dot configurations are established analogously. }
	\label{fig1}
\end{figure}

Our device architecture consists of an undoped silicon channel (Fig.~\ref{fig1}a, dark grey) connected to a highly-doped source (S) and drain (D) reservoir. Metallic polysilicon gates (light grey) partially overlap the channel, each capable of inducing one quantum dot with a controllable number of electrons~\cite{Barraud2016,Houtin2019}. 
While devices with a larger number of split-gate pairs are possible~\cite{Chanrion2020},
we focus on a 2$\times$2 quantum-dot array as the smallest two-dimensional unit cell in this architecture, i.e. a device with two pairs of split-gate electrodes, labelled G$_\mathrm{n}$ with corresponding control voltages $V_\mathrm{n}$. 
The device studied is similar to the one shown in Fig.~\ref{fig1}a, but additionally has a common top gate 300~nm above the channel, and was encapsulated at the foundry by a back-end that includes routing to wirebonding pads. 
Quantum dots are induced in the 7-nm-thick channel by 32-nm-long gates, separated from each other by 32-nm silicon nitride (see Supplementary Information). 
The handle of the silicon-on-insulator wafer is grounded during measurements, but can in principle be utilized as a back gate. 

\begin{figure*}
	\includegraphics[scale=1]{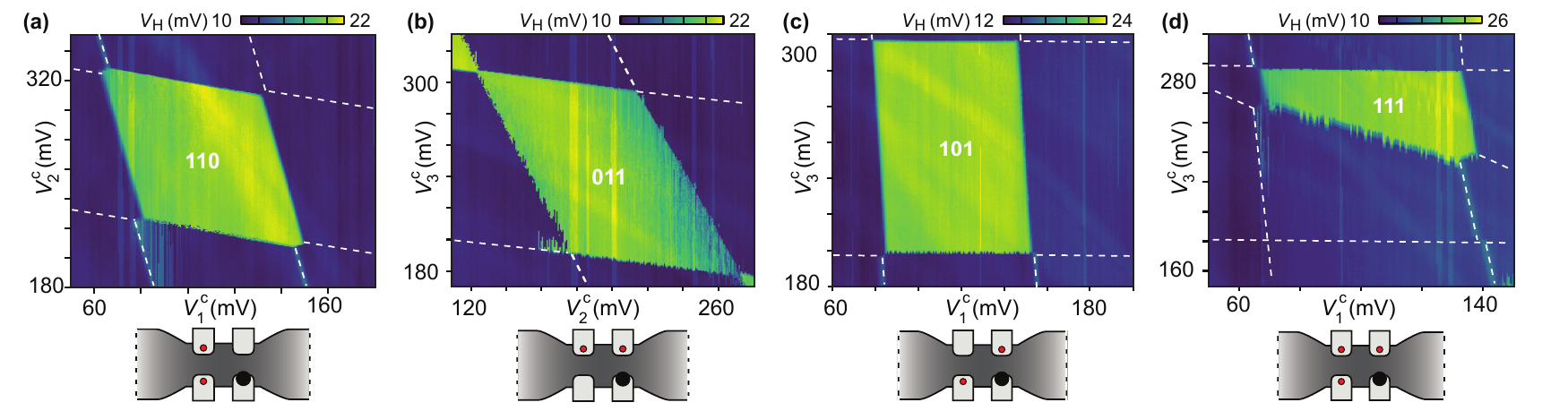}
	\caption{\textbf{Various single-electron configurations within the array.}
		(a)-(c) Three different double dots within the three qubit dots, controlled by compensated voltages $V_\mathrm{1,2,3}^c$. Numbers indicate the occupation of the qubit dots (each red dot represents one electron). (d) Similar to (c), but with $V_\mathrm{2}^c$ fixed at a larger positive voltage, revealing the triple-dot ground-state region.  
In (a)-(d) the top gate is fixed at 6~V.}
	\label{fig2}
\end{figure*}

Figure~\ref{fig1}b shows a schematic of the device with $V_\mathrm{n}$ tuned to induce a few-electron double quantum dot underneath G$_\mathrm{1}$ and G$_\mathrm{4}$. 
Source and drain contacts allow conventional $I(V)$ transport characterization, while an inductor (wirebonded to G$_\mathrm{4}$) allows gate-based reflectometry, in which a radio-frequency carrier ($V_\mathrm{RF}$) and a homodyne detection circuit yields a demodulated voltage $V_\mathrm{H}$~\cite{Volk2019a}. 
Bias tees connected to G$_\mathrm{1-3}$ (not shown) allow the application of high-bandwidth voltage signals. 

Measurement of the source-drain current $I$ as a function of $V_\mathrm{1}$  and $V_\mathrm{4}$ reveals a conventional double-dot stability diagram (Fig.~\ref{fig1}c), with bias triangles arising from a finite source bias $V=-3~$mV and co-tunneling ridges indicating substantial tunnel couplings in this few-electron regime (each dot is occupied by 6-9 electrons). 
The characteristic honeycomb pattern is also observed in the demodulated voltage $V_\mathrm{H}$ (Fig.~\ref{fig1}d, acquired simultaneously with Fig.~\ref{fig1}c), and suggests the potential use of G$_\mathrm{4}$ for (dispersively) sensing charge rearrangements (quantum capacitance) anywhere within the 2D array.  
In the following, we keep dot~4 in the few-electron regime (6-9 electrons, serving as a sensor dot), resulting in an enhancement of $V_\mathrm{H}$ whenever dot~4 exchanges electrons with its reservoir, and reduce the occupation number of the other three dots (which in the single-electron regime we refer to as qubit dots). 
In fact, the large capacitive shift of the dot-4 transition by nearby electrons (evident in Fig.~\ref{fig1}c for dot~1) was used to count the absolute number of electrons within each of the three qubit dots (see Supplementary Information).

It is convenient to control the chemical potential of the three qubit dots without affecting the chemical potential of the sensor dot, as illustrated for dot~1 by the compensated control parameter $V_\mathrm{1}^\mathrm{c}$ (Fig.~\ref{fig1}d). 
This is done experimentally by calibrating the capacitive matrix elements $\alpha_\mathrm{i4}$ such that $V_\mathrm{4}$ compensates for electrostatic cross coupling between $V_\mathrm{1-3}$ and dot 4, i.e. by updating voltage $V_\mathrm{4}=V^\mathrm{o}_\mathrm{4} -\sum_{\mathrm{i}=1}^{3} \alpha_\mathrm{i4} (V_\mathrm{i}-V^\mathrm{o}_\mathrm{i}) $ whenever $V_\mathrm{1-3}$ is changed relative to a chosen reference point $(V^\mathrm{o}_\mathrm{1},V^\mathrm{o}_\mathrm{2},V^\mathrm{o}_\mathrm{3})$. The presence of this compensation is indicated by adding a subscript ``c" to the respective control parameters.  
Using this compensation, and setting the operating point of dot 4 with $V_\mathrm{4}^0$, the associated reflectometry signal  $V_\mathrm{H}$ can be used to detect charge movements between the three qubit dots.

The compensated voltages are used to map out ground-state regions of various desired charge configurations of the qubit array. For example, Figure~\ref{fig2}a was acquired by first parking $V_\mathrm{1}$ and $V_\mathrm{2}$ in the first Coulomb valley of dot~1 and dot~2 (keeping dot 3 empty by setting $V_\mathrm{3}=0$), then tuning $V_\mathrm{4}$ to the degeneracy point of dot~4 (maximum of $V_\mathrm{H}$), before sweeping $V_\mathrm{2}^\mathrm{c}$ vs $V_\mathrm{1}^\mathrm{c}$. The enhancement of $V_\mathrm{H}$ clearly shows the extent of the 110 ground-state region. 
(Here, numbers indicate the occupation of the three qubit dots, as illustrated in the schematics of Fig.~\ref{fig2}.)
Due to the relatively large capacitive coupling of the sensor dot to the qubit dots, dot~4 is in Coulomb blockade outside the 110 region; there $V_\mathrm{H}$ reduces to its approximately constant background. (The gain of the reflectometry circuit had been changed relative to the acquisition in Fig.~\ref{fig1}d). 

In addition to the transverse double dot in Fig.~\ref{fig2}a, we also demonstrate the longitudinal (Fig.~\ref{fig2}b) and diagonal (Fig.~\ref{fig2}c) double dots. 
While such a degree of single-electron charge control is impressive for a reconfigurable, silicon-based multi-dot circuit, it is not obvious how coherent single-spin control (for example via micromagnetic field gradients~\cite{Kawakami2014} or spin-orbit coupling~\cite{Crippa2019}) can most easily be implemented in these foundry-fabricated structures. 
One option is to encode qubits in suitable spin states of 111 triple dots, and operate these as voltage-controlled exchange-only qubits~\cite{Medford2013,Russ2017}. To this end, we demonstrate in Fig.~\ref{fig2}d the tune-up of a 111 triple dot (in order to populate also dot~2, $V_\mathrm{2}^\mathrm{c}$ was chosen more positive relative to Fig.~\ref{fig2}c), revealing the pentagonal boundary expected for the exchange-only qubit~\cite{Gaudreau2006,Medford2013}.

\begin{figure*}
\includegraphics[scale=0.95]{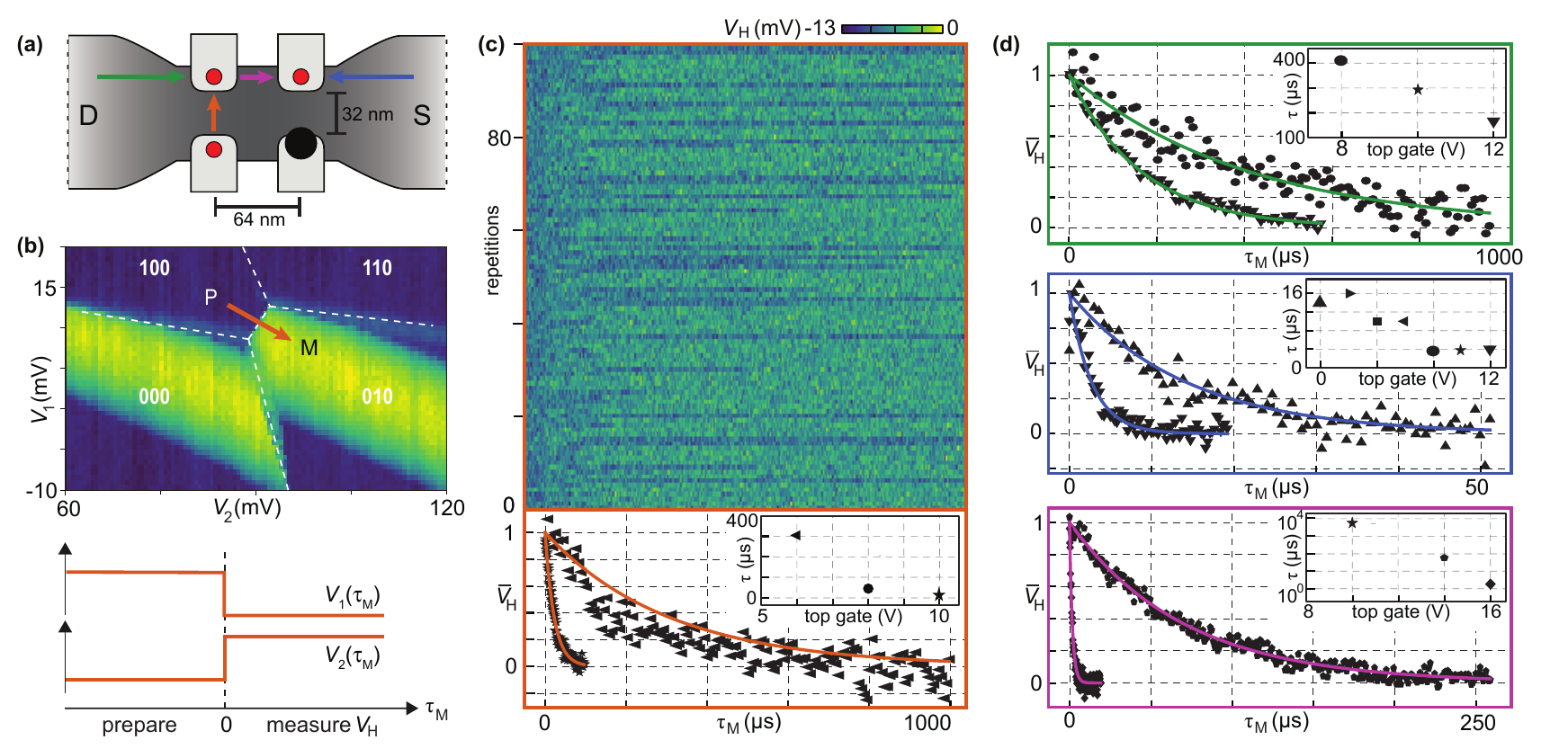}
\caption{
\textbf{Pulsed-gate charge manipulation, single-shot readout, and tunability of tunnel couplings.} 
(a) Device schematic indicating the lead-to-dot (green and blue) and interdot (orange and magenta) transitions for the first electron. 
The arrows indicate the directions of the tunneling events studied. 
(b) Illustration of a $V_\mathrm{1}$-$V_\mathrm{2}$ gate-voltage pulse (orange) that moves an electron from dot-1 to dot-2, with ${V}_\mathrm{4}$ fixed such that a tunneling event causes a change in the sensor signal $V_\mathrm{H}$ (color scale). For each pulse, digitization of $V_\mathrm{H}(\tau_M)$ begins when the gate voltage switches from preparation point P to measurement point M.  
(c) Single-shot traces $V_\mathrm{H}(\tau_M)$ for 100 pulse repetitions, with top gate fixed at 6~V. 
An exponential decay (orange), fitted to the normalized average of all traces ($\blacktriangleleft$), yields a characteristic tunneling time of $300~\mu$s time (inset), and is compared with data obtained with the top gate fixed at 8 ($\bullet$) and 10 V ($\bigstar$). 
(d) Analogously to (c), characteristic trace averages and fitted tunnel times for other single-electron transitions, as color-coded in panel (a).  Insets consistently show a significant decrease of the fitted tunneling times with increasing top-gate voltage.}
\label{fig3}
\end{figure*}

To demonstrate fast single-shot charge readout of the qubit array, we apply voltage pulses to G$_\mathrm{1}$-G$_\mathrm{3}$ while digitizing $V_\mathrm{H}$ \cite{Volk2019a}. 
Specifically, two-level voltage pulses $V_\mathrm{1,2,3}(t)$ are designed to induce one-electron tunneling events into the qubit array or within the array, as illustrated by color-coded arrows in Fig.~\ref{fig3}a. 
One such pulse is exemplified in Fig.~\ref{fig3}b, preparing one electron in dot 1 (P) before moving it to dot 2 (M).
P and M are chosen such that the ground-state transition of interest (in this case the interdot transition) is expected halfway between P and M, using a pulse amplitude of 2~mV.  
This pulse is repeated many times, with $V_\mathrm{4}$ fixed at a voltage that gives good visibility of the transition of interest in $V_\mathrm{H}(\tau_M)$. 
Here, $V_\mathrm{H}(\tau_M)$ serves as a single-shot readout trace that probes for a tunneling event at time $\tau_M$ after the gate voltages are pulsed to the measurement point. 

Figure~\ref{fig3}d shows the repetition of 100 such readout traces obtained at a top gate voltage of 6~V, revealing the stochastic nature of tunneling events, in this case with an averaged tunneling time of~300 $\mu$s. 
This time is obtained by averaging all single-shot traces and fitting an exponential decay. 
In the lower panel of Fig.~\ref{fig3}c, $\bar{V}_\mathrm{H}$ indicates that the average (triangles) has been normalized according to the offset and amplitude fit parameters, which allows comparison with similar data (stars) obtained at a top gate voltage of 10~V (see Supplementary Information). 
The deviation of the data from the fitted exponential decay (solid line) may indicate the presence of multiple relaxation processes, and the reported decay times should therefore be understood as an approximate quantification of characteristic tunneling times within the array. 

While the compact one-gate-per-qubit architecture in accurately-dimensioned Si-MOS devices may ultimately facilitate the wiring fanout of a large-scale quantum computer \cite{Franke2019}, an overall tunability of certain array parameters may initially be essential.
Figure~\ref{fig3}d reports averaged decays for the other transitions within the qubit array, phenomenologically demonstrating that various tunnel couplings within the array can be increased significantly by increasing the common top-gate voltage (insets). 

An important resource for tunnel-coupled two-dimensional qubit arrays is the ability to move or even exchange individual electrons (and their associated spin states) in real space. 
In fact, a two-dimensional triple dot, as in our device, is the smallest array that allows the exchange of two isolated electrons (Heisenberg spin exchange, as demonstrated in linear arrays~\cite{Maune2012}, requires precisely timed wavefunction overlap).

\begin{figure*}
\includegraphics[scale=1]{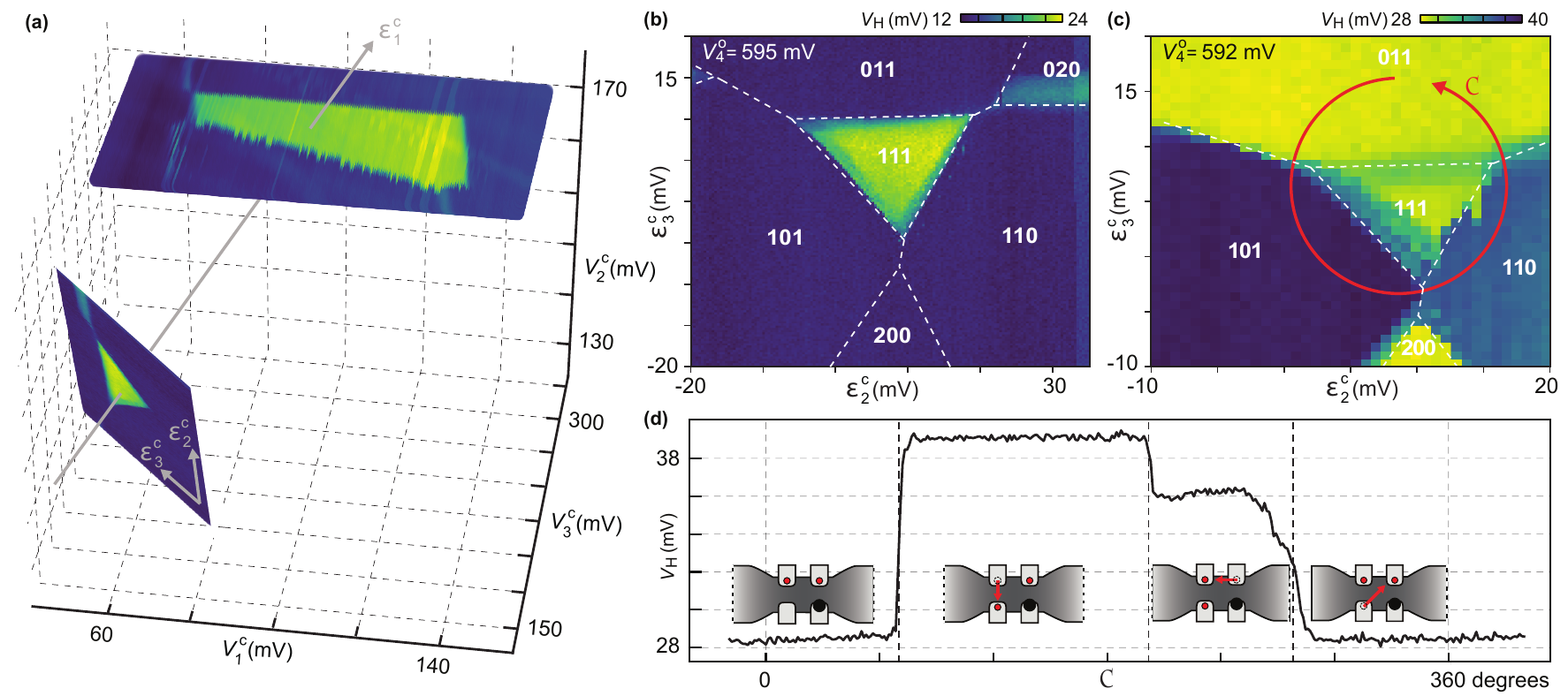}
\caption{\textbf{Physical exchange of two electrons in a 2D array.}
(a) Ground-state region of the 111 triple dot from Fig.~\ref{fig2}d (${V}_\mathrm{2}^c=$constant), plotted in three-dimensional control-voltage space, along measurements within a plane of fixed common mode voltage ($\epsilon_\mathrm{1}^c=$constant). 
Physically, $\epsilon_\mathrm{1}^c$ induces overall gate charge in the array, whereas detuning $\epsilon_\mathrm{2}^c$ ($\epsilon_\mathrm{3}^c$) relocates gate charge within the array along (across) the silicon channel.  
(b) Guides to the eye indicating different ground states within the detuning plane in (a). For this choice of sensor operating point, ${V}_\mathrm{4}^0=595~$mV, ${V}_\mathrm{H}$ does not discriminate between different two-electron configurations. 
(c) Same detuning plane as in (b), but with slightly different sensor operating point, ${V}_\mathrm{4}^0=592~$mV. The control-voltage path $\text{C}$ traverses three two-electron ground states in such a way that the two isolated electrons are exchanged within the array. 
(d) Sensor signal ${V}_\mathrm{H}$ acquired during one cycle of the shuttling path $\text{C}$. Changes in $V_\mathrm{H}$ reflect single-electron movements within the array, as illustrated by red arrows. After completion of one cycle $\text{C}$, the position of the two electrons in the array has been permuted.}
\label{fig4}
\end{figure*}

To demonstrate the spatial exchange of two electrons, we first follow the 111 ground-state region of Fig.~\ref{fig2}d towards lower voltages on G$_\mathrm{1-3}$. 
In Figure~\ref{fig4}a, this is accomplished by reducing the common-mode voltage $\epsilon_\mathrm{1}^\mathrm{c}$, such that the 111 region only borders with two-electron ground states. 
In this gate-voltage region, the qubit array is most intuitively controlled using a symmetry-adopted coordinate system defined by
\begin{equation}
\label{eq:E}	  
\begin{pmatrix}
\epsilon_\mathrm{1}^\mathrm{c}\\
\epsilon_\mathrm{2}^\mathrm{c}\\\
\epsilon_\mathrm{3}^\mathrm{c}
\end{pmatrix} =
\begin{pmatrix}
1/\sqrt{3} & 1/\sqrt{3} & 1/\sqrt{3} \\
0 & -1/\sqrt{2} & 1/\sqrt{2}\\
-2/\sqrt{6} & 1/\sqrt{6} & 1/\sqrt{6}
\end{pmatrix} \begin{pmatrix}
V_\mathrm{1}^\mathrm{c}\\
V_\mathrm{2}^\mathrm{c}\\\
V_\mathrm{3}^\mathrm{c}
\end{pmatrix}.
\nonumber
\end{equation}

Physically, $\epsilon_\mathrm{1}^c$ induces overall gate charge in the qubit array, whereas detuning $\epsilon_\mathrm{2}^c$ ($\epsilon_\mathrm{3}^c$) relocates gate charge within the array along (across) the silicon channel (cf. Fig.~\ref{fig1}b). 
As expected from symmetry, the 111 region within the $\epsilon_\mathrm{2}^c$-$\epsilon_\mathrm{3}^c$ control plane appears as a triangular region, surrounded by the three two-electron configurations 011, 101 and 110, as indicated by guides to the eye in Fig.~\ref{fig4}b.
Importantly, due to the finite mutual charging energies within the array (set by inter-dot capacitances), these three two-electron regions are connected to each other, allowing the cyclic permutation of two electrons without invoking doubly-occupied dots (wavefunction overlap) or exchange with a reservoir.

In principle, any closed control loop traversing 011$\rightarrow$101$\rightarrow$110$\rightarrow$011 should exchange the two electrons, which are isolated at all times by Coulomb blockade, making this a topological operation that may find use in permutational quantum computing~\cite{Jordan2010}. 
In practice, leakage into unwanted qubit configurations (such as 111, 200, 020, etc) can be avoided by mapping out their ground-state regions, as demonstrated in Fig.~\ref{fig4}c by slightly adjusting the operating point ${V}_\mathrm{4}^0$ of the sensor dot. 
This sensor tuning also allows us to verify the sequence of qubit configurations while sweeping gate voltages along the circular shuttling path $\text{C}$, by simultaneously digitizing $V_\mathrm{H}$. The time trace of one shuttling cycle, starting and ending in 011, is plotted in Fig.~\ref{fig4}d, and clearly shows the three charge transitions associated with the two-dimensional exchange (i.e. spatial permutation) of two electrons. 

In this experiment, only gates G1, G2, and G3 can be pulsed quickly, due to our choice of wirebonding G4 as a reflectometry sensor. 
We verified that dot 4 can also be depleted to the last electron (see Supplementary Information), 
and future work will investigate whether the sensor dot can simultaneously serve as a qubit dot. 
Our choice of utilizing dot 4 as a charge sensor (read out dispersively from its gate) realizes a compact architecture for spin-qubit implementations where each gate in principle controls one qubit. 
This technique also alleviates drawbacks associated with the pure dispersive sensing of quantum capacitance, such as tunneling rates constraining the choice of rf carrier frequencies or significantly limiting the visibility of transitions of interest. 
For example, the honeycomb pattern in Fig.~\ref{fig1}d with a clear visibility of dot-4 \emph{and} dot-1 transitions is unusual for gate-based dispersive sensing in the few-electron regime, where small tunneling rates typically limit the visibility of dot-to-lead or interdot transitions~\cite{Gonzalez-Zalba2015}.  
This is a consequence of the strong cross-capacitance between the reflectometry gate G$_\mathrm{4}$ and dot 1, allowing the rf excitation to probe also the quantum capacitances arising from dot 1. 
This also explains the observed discrete features within the bias triangles and shows the potential of gate-based reflectometry for directly revealing excited quantum dot states.
The binary nature of the high-bandwidth charge signal (evident in Fig.~\ref{fig2}) may also simplify the algorithmic tuning of qubit arrays~\cite{Baart2016b}.

While all data presented was obtained at zero magnetic field, application of finite magnetic fields to explore spin dynamics should also be possible~\cite{Maurand2016,Urdampilleta2019}. 
We further expect significant improvements of the reflectometry signal by using superconducting inductors \cite{Ahmed2018} and Josephson parametric amplifiers \cite{Schaal2019}.


In conclusion, we demonstrate a two-dimensional array of quantum dots implemented in a foundry-fabricated silicon nanowire device. 
Each dot can be depleted to the last electron, and pulsed-gate measurements and single-shot charge readout via gate-based reflectometry allow manipulation of individual electrons within the array, while a common top gate provides an overall tunability of tunnel couplings.  
We demonstrate that the array is reconfigurable in situ to realize various multi-dot configurations, and utilize the two-dimensional nature of the array to physically permute the position of two electrons. 
These results constitute key steps towards fault-tolerant quantum computing based on scalable, gate-defined quantum dots. 

\section{Acknowledgements}
We thank Silvano De Franceschi for technical help and the coordination of samples. 
This project received funding from the European Union's Horizon 2020 research and innovation programme under grant agreement MOS-quito (No. 688539). 
F.A. acknowledges support from the Marie Sklodowska-Curie Action Spin-NANO (Grant Agreement No. 676108). 
A.C. acknowledges support from the EPSRC Doctoral Prize Fellowship. 
F.K. acknowledges support from the Independent Research Fund Denmark.

\bibliographystyle{naturemag}

\begin{thebibliography}{10}
\expandafter\ifx\csname url\endcsname\relax
  \def\url#1{\texttt{#1}}\fi
\expandafter\ifx\csname urlprefix\endcsname\relax\def\urlprefix{URL }\fi
\providecommand{\bibinfo}[2]{#2}
\providecommand{\eprint}[2][]{\url{#2}}

\bibitem{Veldhorst2015}
\bibinfo{author}{Veldhorst, M.} \emph{et~al.}
\newblock \bibinfo{title}{{A two-qubit logic gate in silicon}}.
\newblock \emph{\bibinfo{journal}{Nature}} \textbf{\bibinfo{volume}{526}},
  \bibinfo{pages}{410} (\bibinfo{year}{2015}).
\newblock \eprint{1411.5760}.

\bibitem{Muhonen2014}
\bibinfo{author}{Muhonen, J.~T.} \emph{et~al.}
\newblock \bibinfo{title}{{Storing quantum information for 30 seconds in a
  nanoelectronic device}}.
\newblock \emph{\bibinfo{journal}{Nature Nanotechnology}}
  \textbf{\bibinfo{volume}{9}}, \bibinfo{pages}{986} (\bibinfo{year}{2014}).

\bibitem{Watson2018}
\bibinfo{author}{Watson, T.~F.} \emph{et~al.}
\newblock \bibinfo{title}{{A programmable two-qubit quantum processor in
  silicon}}.
\newblock \emph{\bibinfo{journal}{Nature}} \textbf{\bibinfo{volume}{555}},
  \bibinfo{pages}{633--637} (\bibinfo{year}{2018}).

\bibitem{He2019}
\bibinfo{author}{He, Y.} \emph{et~al.}
\newblock \bibinfo{title}{{A two-qubit gate between phosphorus donor electrons
  in silicon}}.
\newblock \emph{\bibinfo{journal}{Nature}} \textbf{\bibinfo{volume}{571}},
  \bibinfo{pages}{371} (\bibinfo{year}{2019}).

\bibitem{Zajac2018}
\bibinfo{author}{Zajac, D.~M.} \emph{et~al.}
\newblock \bibinfo{title}{{Resonantly driven CNOT gate for electron spins}}.
\newblock \emph{\bibinfo{journal}{Science}} \textbf{\bibinfo{volume}{359}},
  \bibinfo{pages}{439--442} (\bibinfo{year}{2018}).

\bibitem{Yoneda2018}
\bibinfo{author}{Yoneda, J.} \emph{et~al.}
\newblock \bibinfo{title}{{A quantum-dot spin qubit with coherence limited by
  charge noise and fidelity higher than 99.9{\%}}}.
\newblock \emph{\bibinfo{journal}{Nature Nanotechnology}}
  \textbf{\bibinfo{volume}{13}}, \bibinfo{pages}{102--106}
  (\bibinfo{year}{2018}).

\bibitem{Mortemousque2018}
\bibinfo{author}{Mortemousque, P.~A.} \emph{et~al.}
\newblock \bibinfo{title}{{Coherent control of individual electron spins in a
  two dimensional array of quantum dots. }} \eprint{Preprint at
  https://arxiv.org/abs/1808.06180 (2018)}.

\bibitem{Dehollain2019}
\bibinfo{author}{Dehollain, J.~P.} \emph{et~al.}
\newblock \bibinfo{title}{{Nagaoka ferromagnetism observed in a quantum dot
  plaquette. }} \eprint{Nature doi:10.1038/s41586-020-2051-0 (2020)}.

\bibitem{Betz2016}
\bibinfo{author}{Betz, A.~C.} \emph{et~al.}
\newblock \bibinfo{title}{{Reconfigurable quadruple quantum dots in a silicon
  nanowire transistor}}.
\newblock \emph{\bibinfo{journal}{Applied Physics Letters}}
  \textbf{\bibinfo{volume}{108}}, \bibinfo{pages}{203108}
  (\bibinfo{year}{2016}).

\bibitem{Maurand2016}
\bibinfo{author}{Maurand, R.} \emph{et~al.}
\newblock \bibinfo{title}{{A CMOS silicon spin qubit}}.
\newblock \emph{\bibinfo{journal}{Nature Communications}}
  \textbf{\bibinfo{volume}{7}}, \bibinfo{pages}{13575} (\bibinfo{year}{2016}).

\bibitem{Crippa2019}
\bibinfo{author}{Crippa, A.} \emph{et~al.}
\newblock \bibinfo{title}{{Gate-reflectometry dispersive readout and coherent
  control of a spin qubit in silicon}}.
\newblock \emph{\bibinfo{journal}{Nature Communications}}
  \textbf{\bibinfo{volume}{10}}, \bibinfo{pages}{2776} (\bibinfo{year}{2019}).

\bibitem{Urdampilleta2019}
\bibinfo{author}{Urdampilleta, M.} \emph{et~al.}
\newblock \bibinfo{title}{{Gate-based high fidelity spin readout in a CMOS
  device}}.
\newblock \emph{\bibinfo{journal}{Nature Nanotechnology}}
  \textbf{\bibinfo{volume}{14}}, \bibinfo{pages}{737--742}
  (\bibinfo{year}{2019}).

\bibitem{Vandersypen2017}
\bibinfo{author}{Vandersypen, L. M.~K.} \emph{et~al.}
\newblock \bibinfo{title}{{Interfacing spin qubits in quantum dots and donors
  Ñ hot , dense , and coherent}}.
\newblock \emph{\bibinfo{journal}{npj Quantum Information}}
  \textbf{\bibinfo{volume}{3}}, \bibinfo{pages}{34} (\bibinfo{year}{2017}).

\bibitem{Barraud2016}
\bibinfo{author}{Barraud, S.} \emph{et~al.}
\newblock \bibinfo{title}{{Development of a CMOS Route for Electron Pumps to Be
  Used in Quantum Metrology}}.
\newblock \emph{\bibinfo{journal}{Technologies}} \textbf{\bibinfo{volume}{4}},
  \bibinfo{pages}{10} (\bibinfo{year}{2016}).

\bibitem{Houtin2019}
\bibinfo{author}{Hutin, L.} \emph{et~al.}
\newblock \bibinfo{title}{{Gate reflectometry for probing charge and spin
  states in linear Si MOS split-gate arrays. }} 
  \eprint{Preprint at https://arxiv.org/abs/1912.10884 (2019)}.

\bibitem{Chanrion2020}
\bibinfo{author}{Chanrion, E.} \emph{et~al.}
\newblock \bibinfo{title}{{Charge detection in an array of CMOS quantum dots. }} 
\eprint{Preprint at https://arxiv.org/abs/2004.01009 (2020)}.

\bibitem{Volk2019a}
\bibinfo{author}{Volk, C.}, \bibinfo{author}{Chatterjee, A.},
  \bibinfo{author}{Ansaloni, F.}, \bibinfo{author}{Marcus, C.~M.} \&
  \bibinfo{author}{Kuemmeth, F.}
\newblock \bibinfo{title}{{Fast Charge Sensing of Si/SiGe Quantum Dots via a
  High-Frequency Accumulation Gate}}.
\newblock \emph{\bibinfo{journal}{Nano Letters}}  (\bibinfo{year}{2019}).

\bibitem{Kawakami2014}
\bibinfo{author}{Kawakami, E.} \emph{et~al.}
\newblock \bibinfo{title}{{Electrical control of a long-lived spin qubit in a
  Si/SiGe quantum dot}}.
\newblock \emph{\bibinfo{journal}{Nature Nanotechnology}}
  \textbf{\bibinfo{volume}{9}}, \bibinfo{pages}{666} (\bibinfo{year}{2014}).

\bibitem{Medford2013}
\bibinfo{author}{Medford, J.} \emph{et~al.}
\newblock \bibinfo{title}{{Self-consistent measurement and state tomography of
  an exchange-only spin qubit}}.
\newblock \emph{\bibinfo{journal}{Nature Nanotechnology}}
  \textbf{\bibinfo{volume}{8}}, \bibinfo{pages}{654} (\bibinfo{year}{2013}).

\bibitem{Russ2017}
\bibinfo{author}{Russ, M.} \& \bibinfo{author}{Burkard, G.}
\newblock \bibinfo{title}{{Three-electron spin qubits}}.
\newblock \emph{\bibinfo{journal}{Journal of Physics: Condensed Matter}}
  \textbf{\bibinfo{volume}{29}}, \bibinfo{pages}{393001}
  (\bibinfo{year}{2017}).
\newblock \eprint{1611.04945}.

\bibitem{Gaudreau2006}
\bibinfo{author}{Gaudreau, L.} \emph{et~al.}
\newblock \bibinfo{title}{{Stability Diagram of a Few-Electron Triple Dot}}.
\newblock \emph{\bibinfo{journal}{Physical Review Letters}}
  \textbf{\bibinfo{volume}{97}}, \bibinfo{pages}{036807}
  (\bibinfo{year}{2006}).

\bibitem{Franke2019}
\bibinfo{author}{Franke, D.~P.}, \bibinfo{author}{Clarke, J.~S.},
  \bibinfo{author}{Vandersypen, L. M.~K.} \& \bibinfo{author}{Veldhorst, M.}
\newblock \bibinfo{title}{{Rent{\textquoteright}s rule and extensibility in
  quantum computing}}.
\newblock \emph{\bibinfo{journal}{Microprocessors and Microsystems}}
  \textbf{\bibinfo{volume}{67}}, \bibinfo{pages}{1--7} (\bibinfo{year}{2019}).

\bibitem{Maune2012}
\bibinfo{author}{Maune, B.~M.} \emph{et~al.}
\newblock \bibinfo{title}{{Coherent singlet-triplet oscillations in a
  silicon-based double quantum dot}}.
\newblock \emph{\bibinfo{journal}{Nature}} \textbf{\bibinfo{volume}{481}},
  \bibinfo{pages}{344--347} (\bibinfo{year}{2012}).

\bibitem{Jordan2010}
\bibinfo{author}{Jordan, S.~P.}
\newblock \bibinfo{title}{{Permutational Quantum Computing}}.
\newblock \emph{\bibinfo{journal}{Quantum Information and Computation}}
  \textbf{\bibinfo{volume}{10}}, \bibinfo{pages}{470} (\bibinfo{year}{2010}).

\bibitem{Gonzalez-Zalba2015}
\bibinfo{author}{Gonzalez-Zalba, M.~F.}, \bibinfo{author}{Barraud, S.},
  \bibinfo{author}{Ferguson, A.~J.} \& \bibinfo{author}{Betz, A.~C.}
\newblock \bibinfo{title}{{Probing the limits of gate-based charge sensing}}.
\newblock \emph{\bibinfo{journal}{Nature Communications}}
  \textbf{\bibinfo{volume}{2}}, \bibinfo{pages}{1--8} (\bibinfo{year}{2015}).

\bibitem{Baart2016b}
\bibinfo{author}{Baart, T.~A.}, \bibinfo{author}{Eendebak, P.~T.},
  \bibinfo{author}{Reichl, C.}, \bibinfo{author}{Wegscheider, W.} \&
  \bibinfo{author}{Vandersypen, L. M.~K.}
\newblock \bibinfo{title}{{Computer-automated tuning of semiconductor double
  quantum dots into the single-electron regime}}.
\newblock \emph{\bibinfo{journal}{Applied Physics Letters}}
  \textbf{\bibinfo{volume}{108}}, \bibinfo{pages}{213104}
  (\bibinfo{year}{2016}).

\bibitem{Ahmed2018}
\bibinfo{author}{Ahmed, I.} \emph{et~al.}
\newblock \bibinfo{title}{{Radio-Frequency Capacitive Gate-Based Sensing}}.
\newblock \emph{\bibinfo{journal}{Physical Review Applied}}
  \textbf{\bibinfo{volume}{10}}, \bibinfo{pages}{014018}
  (\bibinfo{year}{2018}).

\bibitem{Schaal2019}
\bibinfo{author}{Schaal, S.} \emph{et~al.}
\newblock \bibinfo{title}{{Fast Gate-Based Readout of Silicon Quantum Dots
  Using Josephson Parametric Amplification}}.
\newblock \emph{\bibinfo{journal}{Physical Review Letters}}
  \textbf{\bibinfo{volume}{124}}, \bibinfo{pages}{067701}
  (\bibinfo{year}{2020}).

\bibitem{QDevil}
 \eprint{Electronic access: https://www.qdevil.com}.

\end{thebibliography}

\onecolumngrid

\onecolumngrid

\pagebreak
\widetext
\section{Supplementary Information}	
\setcounter{equation}{0}
\setcounter{figure}{0}
\setcounter{table}{0}
\setcounter{page}{1}
\makeatletter
\renewcommand{\theequation}{S\arabic{equation}}
\renewcommand{\thefigure}{S\arabic{figure}}

\subsection{Sample fabrication}
Our quantum-dot arrays are fabricated at CEA-LETI using a top-down fabrication process on 300-mm silicon-on-insolator (SOI) wafers, adapted from a commercial fully-depleted SOI (FD-SOI) transistor technology~\cite{Barraud2016}.
Compared to single-gate transistors (in which a single gate electrode wraps across a silicon nanowire) two main changes in regards to gate patterning are needed in order to realize 2xN arrays. 
First, N gate electrodes are patterned, in series along one silicon channel. Second, a dedicated etching process is introduced that creates a narrow trench through the gate electrodes, along the nanowire, thereby splitting each gate electrode into one split-gate pair~\cite{Houtin2019}. 
The main fabrication steps are described below. 
For illustrative purposes, the device shown in Fig.~\ref{fig1}a was imaged after gate patterning and first spacer deposition~\cite{Barraud2016}, and does not represent the top gate and back-end. 

Starting with a blank SOI wafer (12~nm Si / 145~nm SiO$_2$), the active mesa patterning is performed in order to define a thin, undoped nanowire via a combination of deep-ultra-violet (DUV) lithography and chemical etching.
The silicon nanowire is 7-nm thin after oxidation, and has a width of approximately 70~nm for the device studied in this work. 
Then, a high-quality 6-nm thick SiO$_2$ gate oxide is deposited via thermal oxidation.
To define the metal gate, a 5-nm thick layer of TiN followed by 50~nm of n+ doped polysilicon is used from the standard FD-SOI processing.
The gate is patterned using a combination of conventional DUV lithography combined with an electron-beam lithography process, allowing to achieve an aggressive intergate pitch down to 64~nm (gate length, longitudinal gate spacing, and transverse gate spacing as small as 32~nm). 
Then, 35-nm thick SiN spacers between gates and between gates and source/drain (S/D) regions are formed, which serve two roles:
They protect the intergate regions from self-aligned doping (therefore keeping the channel undoped), and they define tunnel barriers within the array.
Afterwards, raised S/D contacts are regrown to 18~nm to reduce access resistance.
Then, to obtain low access resistances, S/D are doped in two steps: first with lightly-doped drain (LDD) implant (using As at moderate doping conditions) and consecutive annealing to activate dopants, and then with highly-doped drain (HDD) implant (As and P at heavy doping conditions). 
To complete the device fabrication, the gate and lead contact surfaces are metalized to form NiPtSi (salicidation), in preparation for metal lines to be routed to bonding pads on the surface of the wafer. 
Finally, a standard copper-based back-end-of-line process is used to define an optional metallic top gate 300~nm above the nanowire, to make interconnections to bonding pads, as well as to encapsulate the device in a protective glass of siliconoxide.	
Using the powerful parallelism of top-down fabrication, we obtain dozens of dies on a single 300-mm-diameter wafer, each of them containing hundreds of quantum-dot devices buried  2-3~$\mu$m below the chip surface. 

\subsection{Voltage control}

Low-frequency control voltages are generated by a multi-channel digital-to-analog converter (QDevil QDAC)~\cite{QDevil}, whereas high-frequency control voltages are generated using a Tektronix AWG5014C arbitrary waveform generator.
To acquire voltage scans that involve compensated control voltages, we use appropriately programmed QDevil QDACs. 


\subsection{Radio-frequency reflectometry}
The reflectometry technique is similar to that described in Ref. \cite{Volk2019a}, in which a sensor dot tunnel-coupled to two reservoirs was monitored via a SMD-based tank circuit wirebonded to the accumulation gate of the sensor. In this work, the sensor dot (located underneath G$_4$) is tunnel-coupled only to one reservoir (source in Fig.~\ref{fig1}a), and the increased cross-capacitance to the three qubit dots results in much larger electrostatic shifts of dot 4 whenever the occupation of the qubit dots changes. For example, each pair of triple points in Fig.~\ref{fig1}d is spaced significantly larger than the peak width associated with the sensor-dot transition. 

In order to increase the signal intensity as well as to allow for inaccuracies in $\alpha_\mathrm{i4}$, we find it useful to occupy the sensor dot with several electrons (6-9 in Fig.~\ref{fig2}), and to intentionally power-broaden the Coulomb peaks of dot 4 (with -70~dBm applied to the inductor) for all  acquisitions in Fig.~\ref{fig2}.
The SMD inductance used is 820~nH, and the rf carrier has a frequency of a 191.3~MHz. 

For the real-time detection of interdot tunneling events in Fig.~\ref{fig3}c, an Alazar digitizing card (ATS9360) is used with a sample rate set to 500~kS/s. 
The integration time per pixel is set by a 30~kHz low-pass filter (SR560), yielding a signal-to-noise ratio as high as 1.4 in this device.  

\begin{figure}
	\includegraphics[scale=0.12]{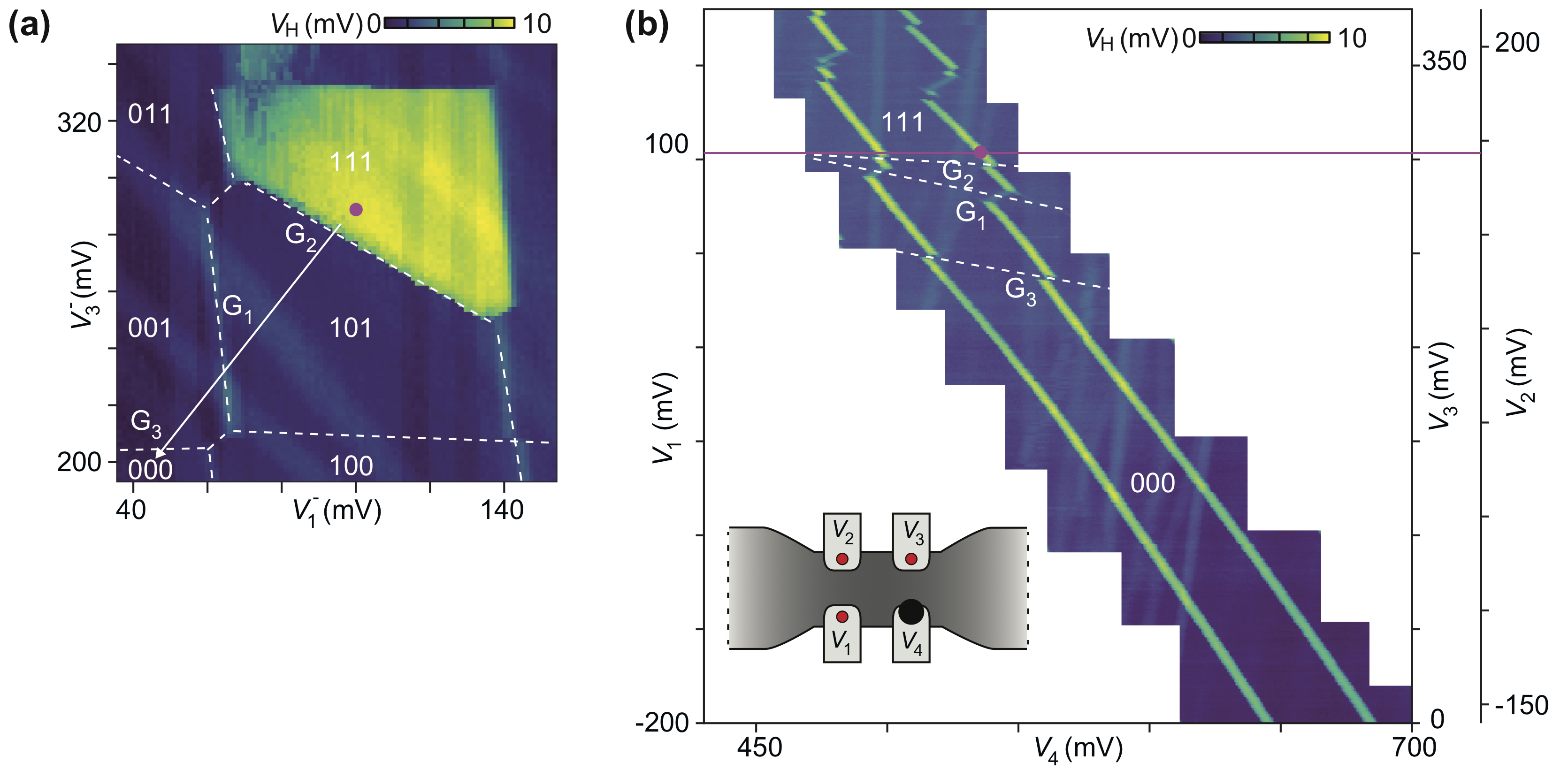}
	\caption{\textbf{Verifying electron count in the 111 configuration.}
(a) Putative 111 ground-state region (yellow area) appearing in a compensated 2D map ($V_1^-$ vs $V_3^-$) similar to the maps in Figure 2. The qubit array, initialized at the magenta dot, can be depleted of electrons by reducing simultaneously $V_1^-$ and $V_3^-$ (white arrow), in this example removing one electron from dot 2, then one from dot 1, and finally one from dot 3 (ground-state transitions are marked by G$_2$, G$_1$, G$_3$).  
However, these charge transitions are not visible in the sensor signal $V_\mathrm{H}$, as dot 4 is in Coulomb blockade (blue regions), thereby neither confirming nor disputing the absolute occupation numbers. 
(b) However, by reducing $V_1$, $V_2$, $V_3$ simultaneously (note the three vertical axes) while sweeping $V_4$, each charge transition within the qubit array induces a capacitive shift of the dot-4 Coulomb peak (magenta dot), confirming the presence of exactly three electrons in the 111 configuration. 
Moreover, comparison with another dot-4 Coulomb peak reveals three distinct slopes (marked G$_2$, G$_1$, G$_3$), confirming that each charge transition indeed corresponds to a different dot. Note the absence of qubit transitions for lower or even negative gate voltages applied to $V_1$, $V_2$, and $V_3$. }
	\label{figs1}x
\end{figure}

\subsection{Determination of electron number}

For a given tuning of the qubit array, the occupation number of each qubit dot is determined by counting the number of discrete electrostatic shifts of the sensor dot (i.e. shifts of a dot-4 Coulomb peak in $V_\mathrm{H}$ along $V_4$) as the qubit array is emptied by continuously reducing the control voltage of the dot of interest. 
If the total number of electrons within the qubit array is desired, all three qubit voltages can be reduced simultaneously, while sweeping $V_4$ over one or more Coulomb peaks of dot 4, which serves as an electrometer. 
An example of such a diagnostic scan, for the case of a 111-occupied triple dot, is shown in Figure~\ref{figs1}. 
To determine the number of electrons in the sensor (dot 4), we utilized Coulomb peaks associated with dot 1 as an electrometer for dot 4, while continuously reducing $V_4$. 
This works because the strong dispersive signal associated with the dot-1-to-lead transition shows discrete shifts (along $V_1$) whenever the dot-4 occupation changes (note the large shifts of the dot-1 transition induced by dot 4 in Fig.~\ref{fig1}d). 

\subsection{Capacitance matrix}

To support our interpretation of dot i being localized predominantly underneath gate i (i=1...4), we extract from stability diagrams the capacitances $C_\mathrm{ij}$ between gate j and dot i (in units of aF):
\begin{equation}
\label{eq:Capmatr}	 \hat{C} =
\begin{pmatrix}
2.14 & 0.33 & 0.25 & 0.73 \\
0.3 & 1.69 & 0.22 & 0.17 \\
0.32 & 0.6 & 1.41 & 0.26 \\
0.79 & 0.34 & 0.47 & 2.00 \\
\end{pmatrix} 
\nonumber
\end{equation}
In this capacitance matrix, the relatively large diagonal elements reflect the strong coupling between each gate and the dot located underneath it. 
By adding several electrons to the array, we have also observed that the capacitances change somewhat, indicating a spatial change of wavefunctions (not shown) and suggesting an alternative way to change tunnel couplings. 

\subsection{Fitting tunneling times}
In Figure~\ref{fig3}c we show 100 single-shot traces (upper panel) and the average of all traces.
The average has been fitted by an exponential decay with the initial value, the 1/e time, and the long-time limit (offset) as free fit parameters. 
For plotting purposes, $\bar{V}_\mathrm{H}$ is then calculated by substracting the offset from the average, and dividing the result by the initial value. 
For clarity of presentation (the sampling rate for raw data of Fig.~\ref{fig3}c was 500kS/s), in the lower panel of Fig.~\ref{fig3}c we also decimated the time bins by a factor of 4. 
	
\clearpage

\end{document}